\documentclass{article}
\usepackage{graphicx,epsfig} 
\usepackage{amsfonts}
\usepackage{amsthm, amsmath}
\usepackage{xcolor}
\usepackage{textcomp}
\usepackage{listings}
\usepackage{soul}
\usepackage{multirow}
\usepackage{makecell}
\usepackage{booktabs}
\usepackage{rotating}
\usepackage[utf8]{inputenc}

\title{Deep Gamma Hedging}
\author{John Armstrong \and George Tatlow}
\date{August 2024}

\begin{document}

\maketitle

\begin{abstract}
We train neural networks to learn optimal replication strategies for an option when two replicating instruments are
available, namely the underlying and a hedging option. If the price of the hedging option matches that of the Black--Scholes
model then we find the network will successfully learn the Black-Scholes gamma hedging strategy, even if the dynamics of the underlying
do not match the Black--Scholes model, so long as we choose a loss function that rewards coping with model uncertainty. Our results
suggest that the reason gamma hedging is used in practice is to account for model uncertainty rather than to reduce the impact
of transaction costs.
\end{abstract}

\section*{Introduction}

In the gamma-hedging strategy one seeks to replicate a derivative
by investing in two or more replicating instruments, typically the underlying and exchange-traded derivatives. The strategy is to ensure that the delta and gamma of the replicating portfolio match the delta and gamma of the derivative at all times. In the classical Black--Scholes model, this strategy would be overkill as continuous-time delta hedging is already sufficient to replicate a derivative. Nevertheless the
gamma-hedging strategy is widely used in practice and a number of heuristic arguments have been put forward to explain its success.

It has long been known that the discrete-time gamma-hedging strategy has a higher order rate of convergence than the delta-hedging strategy. For smooth payoffs, this can be proved using the results on higher-order numerical schemes in \cite{kloedenPlaten}. \cite{gobet2012tracking} give non-sharp lower bounds on the convergence rate on different trading grids even for non-smooth payoffs and \cite{broden2009convergence} improves this result to obtain convergence of order $N^{-\frac{3}{4}}$.
If the hedging-strategy converges more rapidly as the re-hedging interval tends to zero, one might expect that by allowing for less frequent rebalancing the gamma-hedging strategy might reduce transaction costs.

More recently \cite{armstrong2023gamma} provides an alternative
motivation for pursuing the gamma-hedging strategy. If one can predict how option prices will change in response to changes in the underlying, then the gamma-hedging
strategy converges irrespective the path of the underlying, so long as it is sufficiently regular. By contrast, delta-hedging converges only if the underlying has quadratic variation matching a particular probability model. We see that delta-hedging arises if one assumes a probabilistic model for the underlying, whereas gamma-hedging arises from a non-probabilistic model of how option prices respond to changes in the underlying.

This provides two new possible motivations for gamma-hedging. First, one might pursue gamma-hedging because one finds it easier to predict the response of the market to a change than it is to model the underlying. This would seem to underpin the market practice of calibration where one fits a pricing model to market prices rather than to historical trends in the underlying. Second, the convergence results of \cite{armstrong2023gamma} are based on rough-path theory which has better continuity properties than classical It\^o calculus. As a result, one might pursue gamma-hedging to add robustness to a classical delta-hedging strategy.

\medskip

The current paper seeks to explore the potential benefits of gamma-hedging using deep learning. Using deep-learning we can calculate good approximations of optimal trading strategies in the face of complexities such as transaction costs. Classical analytical approaches can only examine such issues asymptotically. By examining the strategies found using deep-learning in the face of such complexities we can determine under what conditions gamma-hedging will emerge as a close to optimal strategy. In particular we will be able to compare the relative importance of addressing transaction costs and model robustness as motivations for gamma hedging. Hence, this paper seeks to use deep learning primarily as a tool
to develop an explanatory model for the known market practice
of gamma hedging.

Nevertheless, our approach also shows that if one wishes to develop optimal trading strategies using deep-learning then there may be considerable benefits to be obtained by considering trading strategies in instruments beyond the underlying. Thus our paper also contributes to the growing literature on the use of deep-hedging to develop optimal trading strategies.

The first paper which proposed applying neural networks to the problem of hedging derivatives was \cite{RePEc:nbr:nberwo:4718} in 1994, however, it was \cite{bühler2018deep} in 2018 which actually demonstrated a modern deep-reinforcement learning framework for dynamically hedging a portfolio of derivatives in the presence of market frictions and coined the term \lq Deep Hedging'. Since then, the team behind \cite{bühler2018deep}, have continued to advance the topic in collaboration with others.  For instance, \cite{wiese2019deephedginglearningsimulate} addresses the lack of useful real market data for training a neural network in most hedging scenarios, by focusing on simulating realistic synthetic market data for option prices. \cite{horvath2021deephedgingroughvolatility} investigates the performance of the deep-hedging framework in the presence of rough volatility models. In \cite{murray2022deephedgingcontinuousreinforcement}, they introduce an approach to training a single network to derive the optimal hedging strategy for various portfolios and levels of risk aversion simultaneously. This means that if you want to adjust your risk preferences or if the composition of the portfolio you're hedging changes, you would not need to retrain the model. More recently, they have even looked at the potential of using quantum neural networks when deep hedging in \cite{Cherrat2023quantumdeephedging}.

The paper \cite{bühler2018deep}  also inspired an increased interest from others in applying deep learning to the problem of hedging financial products. \cite{carbonneau2020deephedginglongtermfinancial} extended the algorithm introduced by Bühler et al. to focus on deep hedging long-term financial derivatives, while \cite{Cao_2020} further investigated the ability of deep hedging to find an optimal strategy in the presence of transaction costs. Another notable paper is \cite{Gao_2023}, which also focuses on training deep hedging neural networks using data which better resembles real market dynamics. They achieve this by introducing an improved model for generating synthetic market data which they call the Chiarella-Heston model.

Aside from \cite{carbonneau2020deephedginglongtermfinancial} and \cite{wiese2019deephedginglearningsimulate}, all of the papers mentioned which followed \cite{bühler2018deep} use delta hedging as a benchmark traditional hedging method with which they compare the performance and strategy from their deep hedging model. Our paper contributes the literature by considering the alternative benchmark of gamma-hedging. It also contributes to the literature
on deep-hedging by showing that one can successfully incorporate model uncertainty into the objective function.

As we will see in Section \ref{Result Sec: Method Comparison}, the deep-hedging strategies which use both the underlying plus and an option markedly outperform strategies which only use the underlying. We demonstrate this both when the dynamics of the underlying are known (in which case we use a classical loss function) and when there is model uncertainty (where our loss function incorporates a maximum over different probability models). We find that when there is model uncertainty, the optimal strategy is close to gamma-hedging even when there are transaction costs. Without model uncertainty, diverging significantly from gamma-hedging leads to better results. We consider a variety of different transaction cost models, including no transaction costs. In this case the asymptotically optimal strategy is known to be gamma-hedging and so this provides good evidence that the deep-learning approach is able to identify optimal strategies even when one increases the number of underlyings and incorporates model uncertainty.

\section{Modelling Framework}

We will consider a number of optimization problems related to hedging in the Black-Scholes model but incorporating the additional features of: proportional transaction costs, model uncertainty and using multiple hedging instruments. In this section we will describe a general simulation and optimization model which is able to describe all the phenomena we will study.

We will simulate the optimal behaviour of a trader who reacts to changes in a stock price signal $S_t$ for $t \in [0,T^\prime]$. The trader writes a derivative contract at time $0$ and attempts to replicate the payoff of this derivative by trading in $M$ exchange traded derivatives for times $t \in [0,T^\prime]$.

Because we wish to consider the gamma-hedging strategy on the interval $[0,T^\prime]$ we want the gamma of all these derivative contracts to be finite on the whole of $[0,T^\prime]$. To achieve this we will assume that the derivative contracts are all European options with smooth payoff functions at time $T^\prime$. For computational convenience we will further assume that the payoff functions at time $T^\prime$ are given by the Black-Scholes formula for the price of call options with maturity at time $T$, where $T>T^\prime$. Thus our simulations can be interpreted as considering the replication of standard European call options up to a time $T^\prime$ on the assumption that the portfolio can be liquidated at this time at the Black-Scholes price for all options.

Viewing these derivatives as call options with maturity $T$, We will write $K^\alpha$ for the strike price of option $\alpha$. The index $\alpha=0$ denotes the option we seek to replicate. The indices $1\leq \alpha \leq M$ denote the options we use as hedging instruments.

We will assume that we know how the exchange traded option prices change in response to the stock price signal $S_t$. We will assume that at all times $t \in [0,T^\prime]$ the price of the option with strike $K^\alpha$ is given by the Black-Scholes formula
\begin{equation}
    V^\alpha_t = \mathrm{BS}(S_t, K^\alpha, r, T-t, \sigma).
\end{equation}
Here $\sigma$ is a fixed value. This assumption is intended to model the situation
where a trader has calibrated a model (in this case the Black-Scholes model) to market data and is using this to determine how they believe market option prices
will change in response to the signal.

We do not, however, always assume that the stock price signal $S_t$ has volatility $\sigma$. Instead we will assume that the stock price follows the process
\begin{equation}
    dS_t=S_t(\mu_\eta\,dt + \sigma_\eta\, dW_t), \qquad S_0=S_\eta
\end{equation}
where $\mu_\eta$ and $\sigma_\eta$ are random parameter values and $S_\eta$ is a random initial condition. We will write $(\Omega, {\cal F}, {\mathbb P})$ for the associated probability space.

We will assume that there are proportional transaction costs $p^\alpha$ given when purchasing the derivative $\alpha$. The trader will only trade at evenly spaced discrete time points $\{0, \delta t, 2 \delta t, \ldots, (n-1) \delta t\}$ where $n \delta t=T^\prime$.

The quantities held by the trader at time $t$ are denoted by $q^\alpha(t)$. We assume that $q^\alpha(i \delta t)$ is a deterministic function depending upon the values $(S_{j \delta t})_{0\leq j \leq i})$. In practice $q^\alpha$ will either be computed using a neural network or will be determined by a classical hedging strategy such as delta or gamma hedging.

The profit and loss of the strategy is then given by the following random variable.
\begin{align}\label{PnL}
    \text{PnL}(\omega)=&V^0_0+\sum^M_{\alpha=1}\sum^{n-1}_{i=0}\Bigl(q^\alpha_{i\delta t}(V^\alpha_{(i+1)\delta t}-V^\alpha_{i\delta t})\Bigr) \nonumber \\
    -&V^0_1-\sum^M_{\alpha=1} p^\alpha \Bigl(|q^\alpha_0 \,V^\alpha_0|+\sum^n_{i=1}|V^\alpha_{i\delta t}|\cdot|q^\alpha_{i\delta t}-q^\alpha_{(i-1)\delta t}|\Bigr)
\end{align}
where $\forall \alpha \in \{1,2,...,M\}$, $ q_{T^\prime}^\alpha=0$.

To model the optimal behaviour of the trader we will generate a finite set of samples $\Omega^\prime \subset \Omega$ and will compute the functions $q^\alpha$ using a neural network
which minimizes some loss function. The two loss functions we will consider are:

\begin{equation}\label{loss_abs}
\mathcal{L}^\text{mean}(\Omega^\prime)=\frac{1}{|\Omega^\prime|} \sum_{\omega \in \Omega^\prime} |\text{PnL}(\omega)|
\end{equation}
and
\begin{equation}\label{loss_rob}
    \mathcal{L}^\text{max}(\Omega^\prime)=\max_{\omega \in \Omega^\prime}|\text{PnL}(\omega)|.
\end{equation}

The first loss function will be used for an agent without model uncertainty. The second loss function will be useful to model a trader with model uncertainty who wishes to choose a robust
hedging strategy.

\section{Delta and Gamma Hedging}\label{Section: Delta and Gamma Hedging}

We compare the strategies derived by our deep-learning models with the delta and gamma hedging strategies. The delta and gamma of an option are the first and second derivative of its price with respect to the price of its underlying, $S$, respectively. Under the Black-Scholes option pricing formula, these values are given by
\begin{equation}\label{eq: Delta}
    \Delta = N(d_1)
\end{equation}
\begin{equation}\label{eq: Gamma}
    \Gamma = \frac{N'(d_1)}{S_t \sigma \sqrt{T-t}}
\end{equation}
with,
\begin{equation}
    d_1 = \frac{\ln\left(\frac{S_t}{K}\right) + \left(r + \frac{\sigma^2}{2}\right)(T-t)}{\sigma \sqrt{T-t}}
\end{equation}
where $N(\cdot)$ and $N'(\cdot)$ are the cdf and pdf of a standard normal distribution respectively.

Following the delta-hedging strategy, means choosing your hedging positions to ensure that the overall delta of your portfolio is equal to zero. Gamma hedging requires you to choose a portfolio which has a zero overall gamma, as well as delta.

In section \ref{Result Sec: Method Comparison}, we show how the delta- and gamma-hedging strategies perform over a set of scenarios. In this situation we will have two hedging instruments: Option 1 with $K^1=0$ (equivalent to using the underlying $S$), and Option 2 with $K^2\neq0$.

For the delta hedging strategy we use only Option 1 to hedge. Using (\ref{eq: Delta}) we get $\Delta^\text{1}=1$. Thus, to follow the delta-hedging strategy requires us to take a position of $q_t^{1,\Delta}=\Delta_t^0$ at each timepoint.

For the gamma hedging strategy we must use both options. Since $\Delta^{1}$ is always equal to 1, we know that $\Gamma^{1}=0$, meaning that Option 1 does not effect the gamma of the overall portfolio. Therefore, to make the portfolio gamma neutral we hold $q_t^{2,\Gamma}$ of Option 2 at each timepoint, with
$$q_t^{2,\Gamma}=\frac{\Gamma_t^0}{\Gamma_t^2}$$  We then use Option 1 to make the portfolio delta neutral, holding $q^{1,\Gamma}_t$ at each timepoint, where
$$q_t^{1,\Gamma}=\Delta_t^0 - q_t^2 \Delta_t^2$$

\section{Neural Network Design}

In our calculations where the effects of transaction costs are not considered, we use a single feedforward neural network consisting of three hidden layers, each containing 100 nodes. Within the hidden layers the activation function applied is the ReLU function, whilst in the output layer we use the Identity function.

This network is applied at every timepoint $t \in \{0, \delta t, 2 \delta t, \ldots, (n-1) \delta t\}$. At each point it takes two inputs $(t,S_t)$ and gives M outputs $\{q_t^\alpha\}^M_{\alpha=1}$. The outputs from all timepoints can then be concatenated and used as input for the appropriate loss function.

\begin{figure}[htp!]
\centering
    \includegraphics[width=10cm]{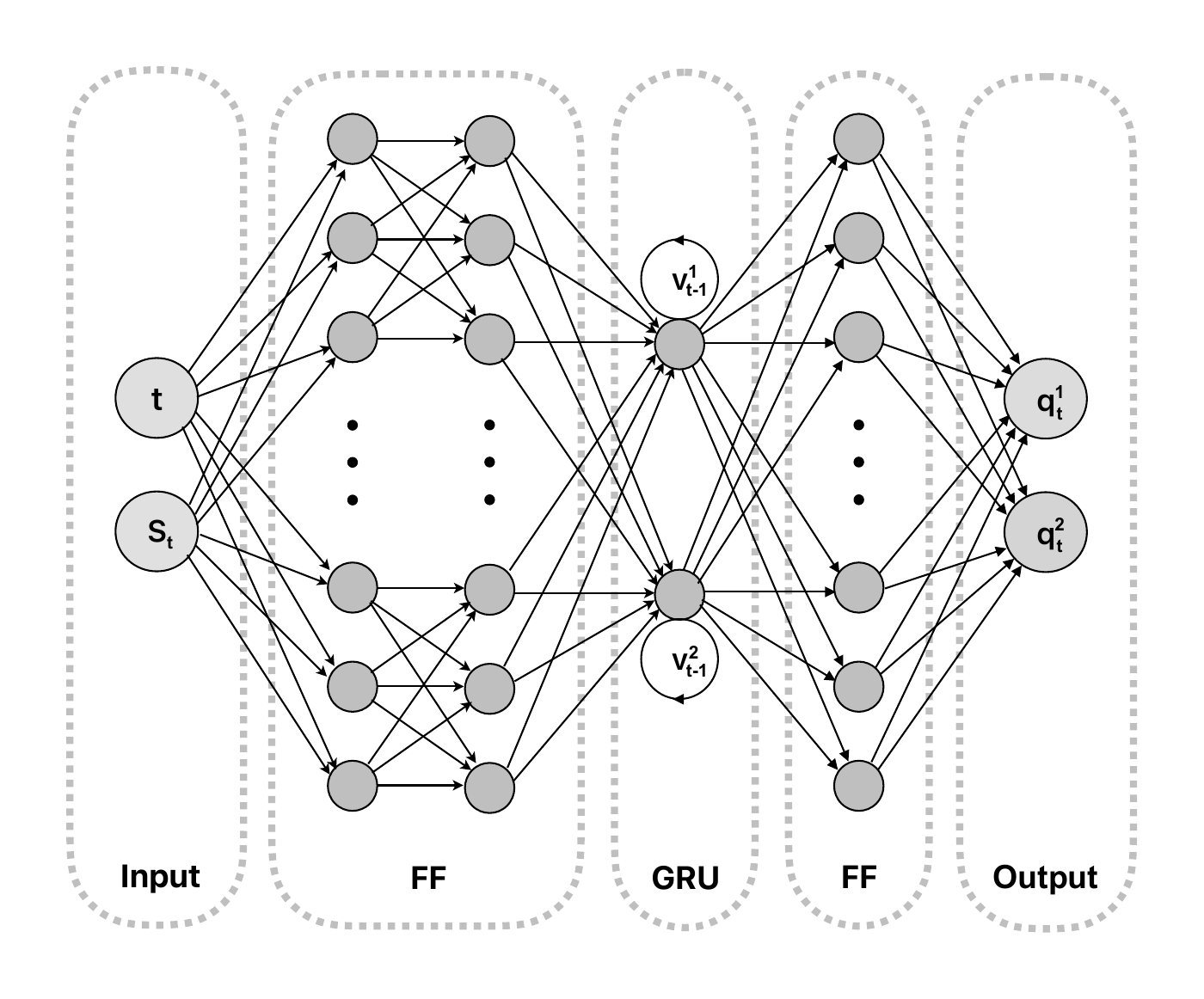}
    \label{fig:NN Drawing}
    \caption{\small{GRU NN architecture}}
\end{figure}

When transaction costs are non-zero, we use a recurrent network, introducing a GRU layer between the second and third hidden layers of the previous structure. We set the number of GRU cells in this layer to be equal to $M$. The reason for this is so that the network has the capacity to pass forward information about the positions taken at the previous timepoint, but is not able to make statistical inferences regarding the volatility or drift for each individual path. Using more cells or LSTM cells would have given it more capacity to do this. Aside from introducing this GRU layer, we keep the the properties of the input, output and the other three hidden layers the same. Thus the new architecture for the case where M=2 can be visualized as in figure \ref{fig:NN Drawing}, where we expect that $v^1_{t-1}$ and $v^2_{t-1}$, which are fed in from the previous time step, will provide information on the values of $q^1_t-\delta t$ and $q^2_t-\delta t$.

All neural networks in this paper are trained using the Adam algorithm from \cite{kingma2017adam} over 20 epochs with a batch size of 256.

\section{Numerical Results}

For the models demonstrated in sections \ref{Result Sec: Lmean- no trans}-\ref{Result Sec: Deep hedging with trans} we set $M=2$, where the two options used to hedge have strike values of $K^1=0$ and $K^2=1.1$. In section \ref{Result Sec: Method Comparison}, we then summarise the performance of all the models in section \ref{Result Sec: Deep hedging with trans}, along with each of their corresponding models with $M=1$, where the only option used to hedge has a strike value of $K^1=0$. 

We used the following common parameter values in all of our simulations: 

\begin{equation*}
    r=0,\,\, \sigma=0.2 ,\,\, T^\prime=1 ,\,\, K^0=1 ,\,\, T=1.4 ,\,\, S_\eta \sim \text{Uniform}(0, 2)
\end{equation*}

In Subsections \ref{Result Sec: Lmean- no trans} and \ref{Result Sec: Lmax- no trans}, where we utilize the feedforward neural network, we set $|\Omega^\prime|=2.56\times 10^6$ . Following this, in Subsections \ref{Result Sec: Deep hedging with trans} and \ref{Result Sec: Method Comparison}, employing the recurrent network and an increased number of timepoints, we reduce the value to $|\Omega^\prime|=2.56\times 10^5$ due to the increased training time required by the network.

We apply the $\mathcal{L}^\text{mean}$ in the cases of no model uncertainty. Therefore, for these simulations we have $\sigma_\eta=\sigma=0.2$ and $\mu_\eta=0$. Under $\mathcal{L}^\text{max}$ hedging, where we admit model uncertainty, we model this using $\sigma_\eta \sim \text{Uniform}(0, 0.3)$ and $\mu_\eta \sim \text{Uniform}(-0.05, 0.1)$.

Our volatility values and our proportional transaction cost values in section \ref{Result Sec: Deep hedging with trans} are based on market data for the S\&P 500\footnote{On October 30th 2023, the spread on the Vanguard S\%P 500 ETF was 0.01\% and on an approximately at the money Call Option expiring on 30th Sept 2024 it was 0.3\%. The VIX opened at 19.75 and its high for the previous year was 30.81. }.

\subsection{$\mathcal{L}^\text{mean}$ hedging without transaction costs}\label{Result Sec: Lmean- no trans}

For our first model we look at hedging using the $\mathcal{L}^\text{mean}$ loss function in the absence of transaction costs and re-hedging over 40 timepoints, ($p^1=p^2=0$, $n=40$).

\begin{sidewaysfigure}[htp!]
\centering
\begin{tabular}{ccc}
\small{$t=0.1$} & \small{$t=0.5$} & \small{$t=0.975$} \\
\includegraphics[width=6cm]{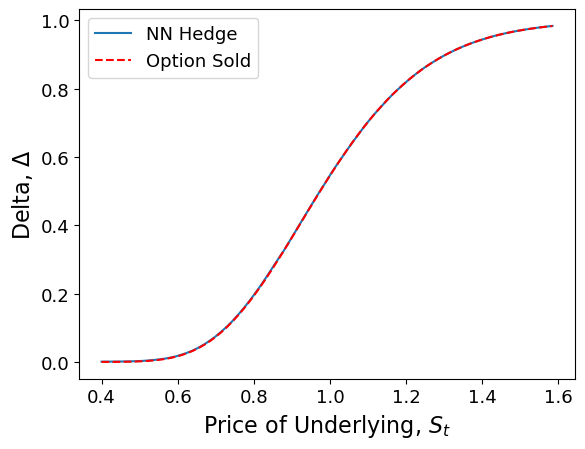} &
\includegraphics[width=6cm]{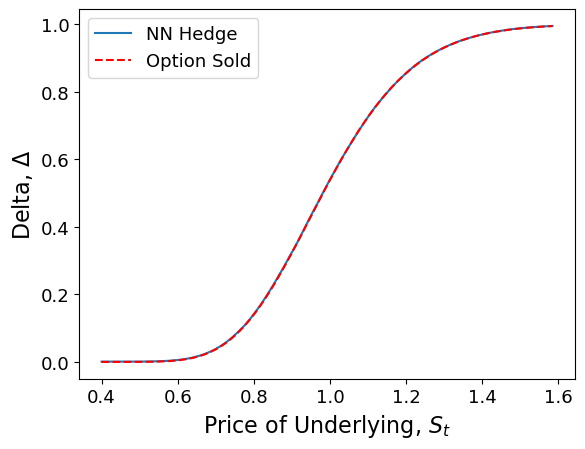} &
\includegraphics[width=6cm]{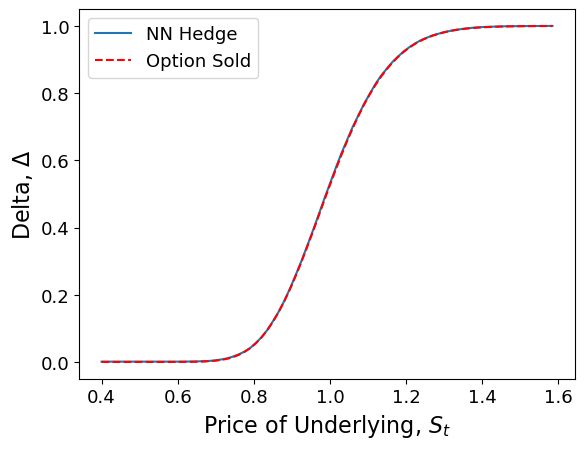} \\
\includegraphics[width=6cm]{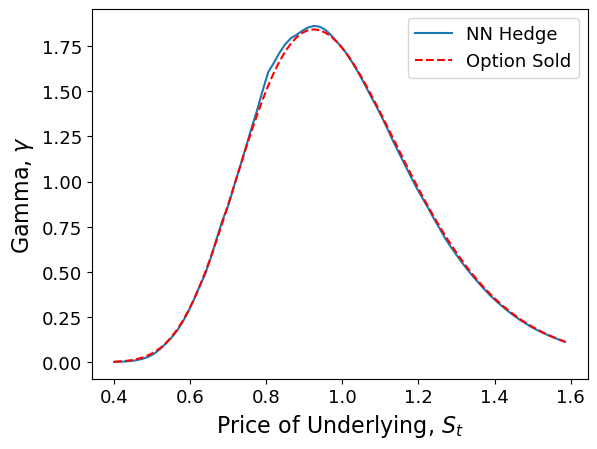} &
\includegraphics[width=6cm]{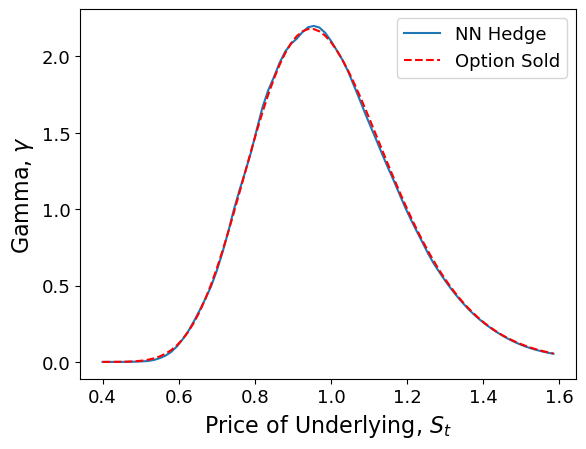} &
\includegraphics[width=6cm]{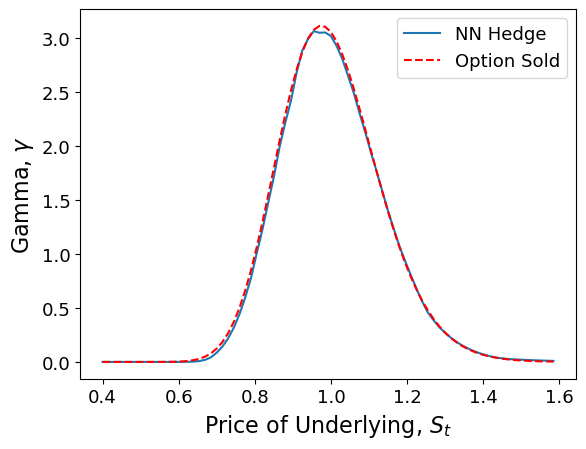} 
\end{tabular}
    \caption{\small{The graphs on the top and bottom row compare the delta and gamma values respectively, of the hedging position given by the neural network with those of the option being hedged.}}
    \label{fig:Pt 1 Model 1 Results}
\end{sidewaysfigure}

In Figure \ref{fig:Pt 1 Model 1 Results} we demonstrate the hedging strategy taken by the network. The graphs on the top row show the total $\Delta$ of the portfolio chosen by the neural network, along with that of Option 0, for varying $S$ with $t$ fixed. The graphs on the bottom row do the same but for the $\Gamma$. We can see that the network chooses a gamma-neutral portfolio at all timepoints.

\graphicspath{{Pt1/Model1}}

\newpage

\subsection{$\mathcal{L}^\text{max}$ hedging without transaction costs}\label{Result Sec: Lmax- no trans}

We now look at hedging using the $\mathcal{L}^\text{max}$ loss function in the absence of transaction costs ($p^1=p^2=0$), but allowing for model uncertainty. Under our given parameters we train three different models, each with a different value for $n$.

\begin{sidewaysfigure}
\begin{tabular}{ccc}
    $n=5$ & $n=10$ & $n=40$ \\
  \includegraphics[width=6cm]{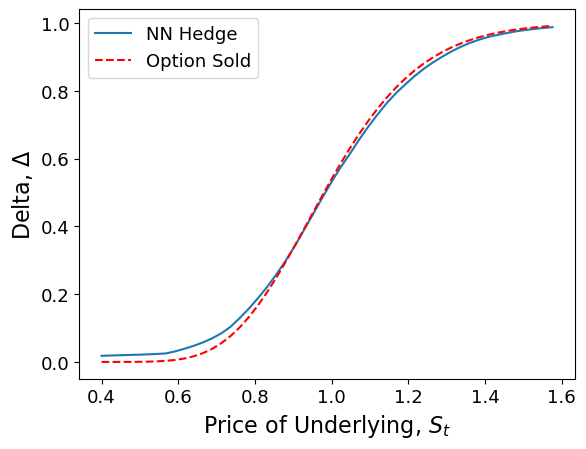} &
  \includegraphics[width=6cm]{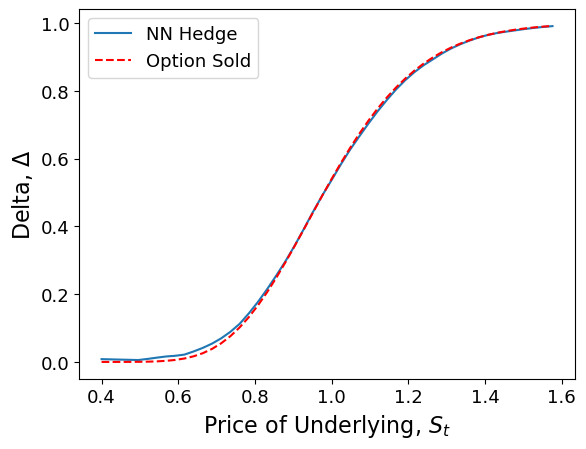} &
  \includegraphics[width=6cm]{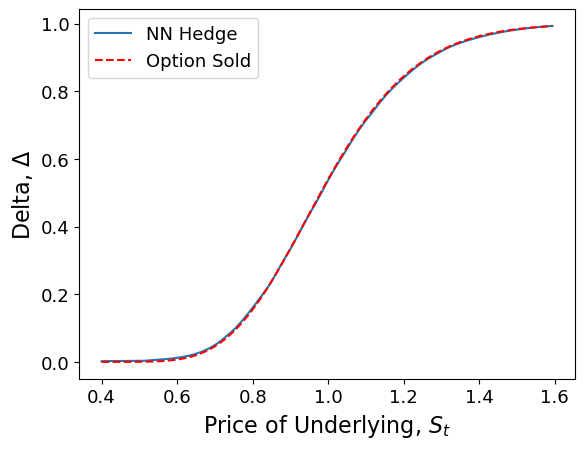} \\
  \includegraphics[width=6cm]{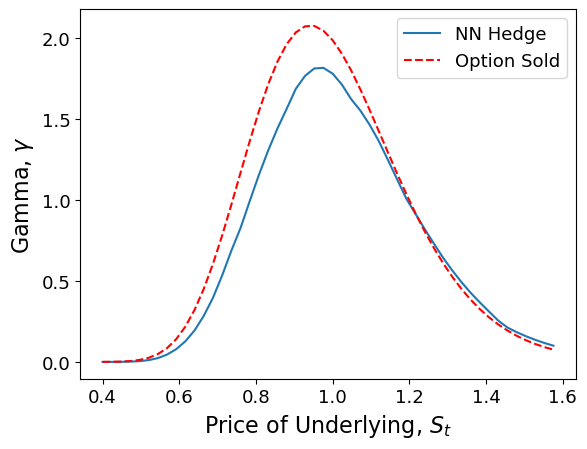} &
  \includegraphics[width=6cm]{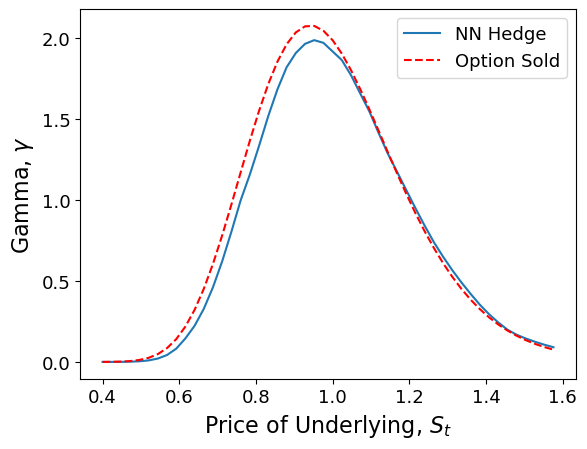} &
  \includegraphics[width=6cm]{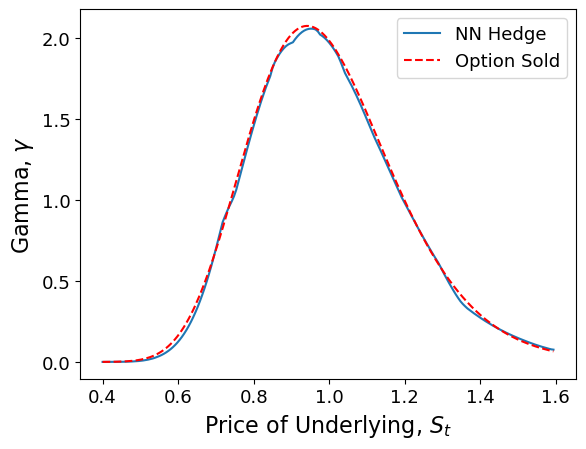} \\
\end{tabular}
    \caption{\small{The graphs on the top and bottom row compare the delta and gamma values respectively, of the hedging position given by the neural network with those of the option being hedged at time t=0.4. Each column uses a NN trained using a different number of timepoints, $n$.}}
    \label{fig:Pt 2 Model 1 Results}
\end{sidewaysfigure}

In Figure \ref{fig:Pt 2 Model 1 Results} we plot the same graphs as in the previous section, but now for different values of $n$ rather than $t$. We can see that as we increase the values of $n$ and thus $\delta t \to 0$, the optimal robust hedging strategy tends to the gamma-hedging strategy.

\graphicspath{{Pt2}}

\subsection{Hedging with transaction costs}\label{Result Sec: Deep hedging with trans}

Now we consider the optimal strategy for both loss functions in the presence of transaction costs. It is at this point that we use the recurrent neural network containing two GRU cells. This allows our network to feed itself information on it's current position before choosing what positions to take next. We also greatly increase the number of time steps to allow for more realistic strategies.

We evaluate the performance of this network under three different sets of proportional transaction costs: no transaction costs ($p^1=p^2=0.0\%$); normal transaction costs ($p^1=0.005\%, p^2=0.25\%$) based on market data for the S\&P 500; and high transaction costs ($p^1=p^2=0.5\%$).

Figures \ref{fig:Abs, Trans - No}, \ref{fig:Abs, Trans - Low} and \ref{fig:Abs, Trans - High} look at the hedging strategy chosen for a single random price path of the underlying, by networks trained under zero, normal and high transaction costs respectively, all using the $\mathcal{L}^\text{mean}$ loss function. Figures \ref{fig:Rob, Trans - No}, \ref{fig:Rob, Trans - Low} and \ref{fig:Rob, Trans - High} do the same but under the $\mathcal{L}^\text{max}$ loss function. In each figure, the graphs on the top show the total delta and gamma of the portfolio chosen by the network at each time alongside that of Option 0. The graphs on the bottom show the positions taken in each of the two hedging options over time.

\graphicspath{{Pt3}}

\begin{figure}[htp!]
\centering
    \includegraphics[width=12cm]{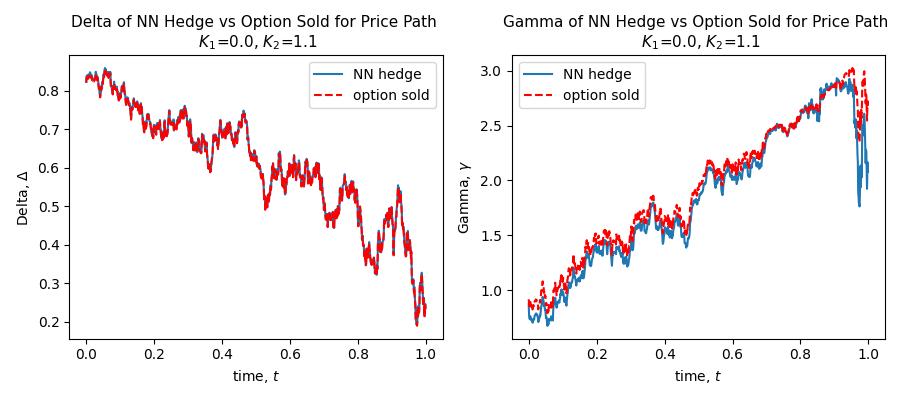}
    \includegraphics[width=12cm]{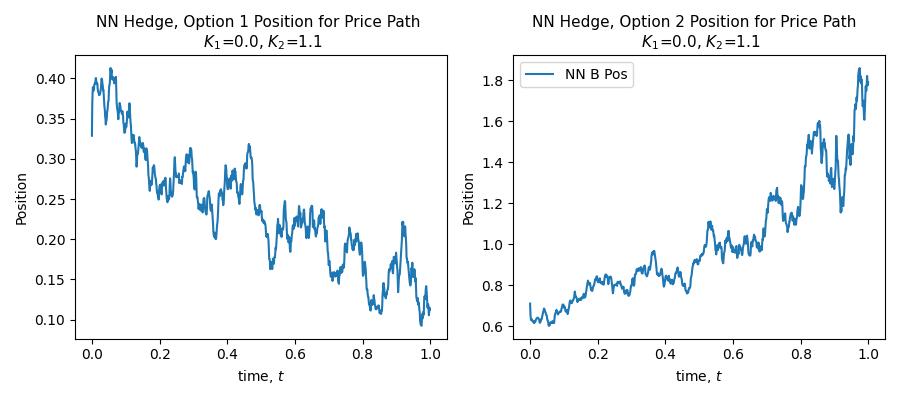}
    \caption{\small{Comparing the gamma hedging method with the strategy chosen by our deep hedging model, applying the $\mathcal{L}^\text{mean}$ loss and no transaction costs.}}
    \label{fig:Abs, Trans - No}
\end{figure}

\begin{figure}[htp!]
\centering
    \includegraphics[width=12cm]{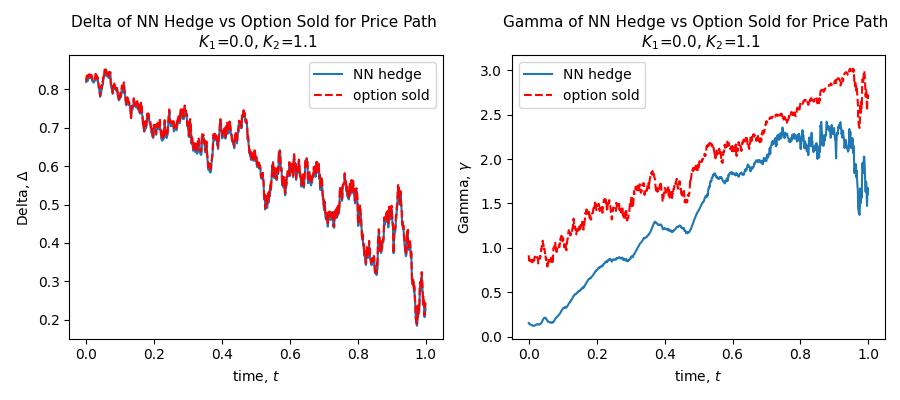}
    \includegraphics[width=12cm]{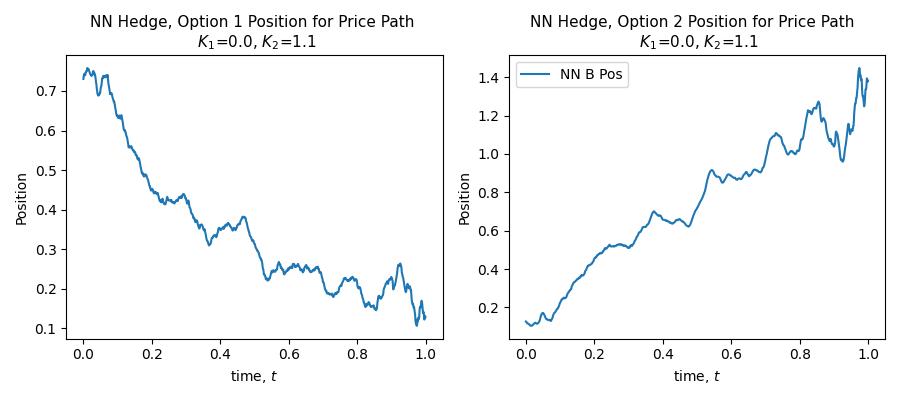}
    \caption{\small{Comparing the gamma hedging method with the strategy chosen by our deep hedging model, applying the $\mathcal{L}^\text{mean}$ loss and normal transaction costs.}}
    \label{fig:Abs, Trans - Low}
\end{figure}

\begin{figure}[htp!]
\centering
    \includegraphics[width=12cm]{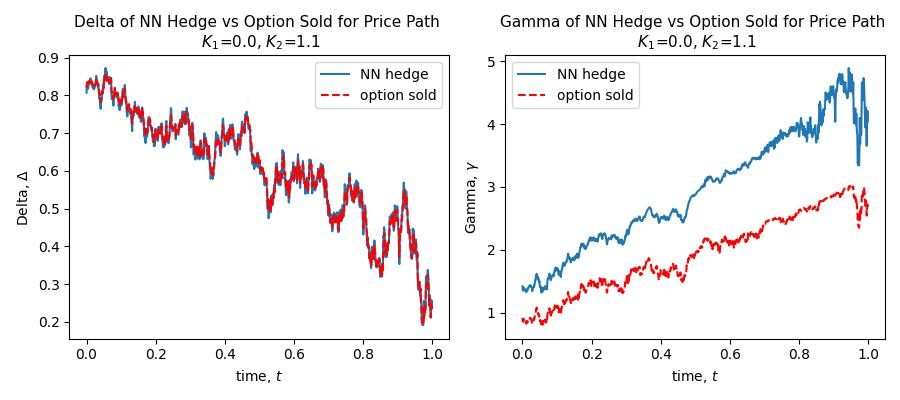}
    \hspace*{-0.9cm}
    \includegraphics[width=12.5cm]{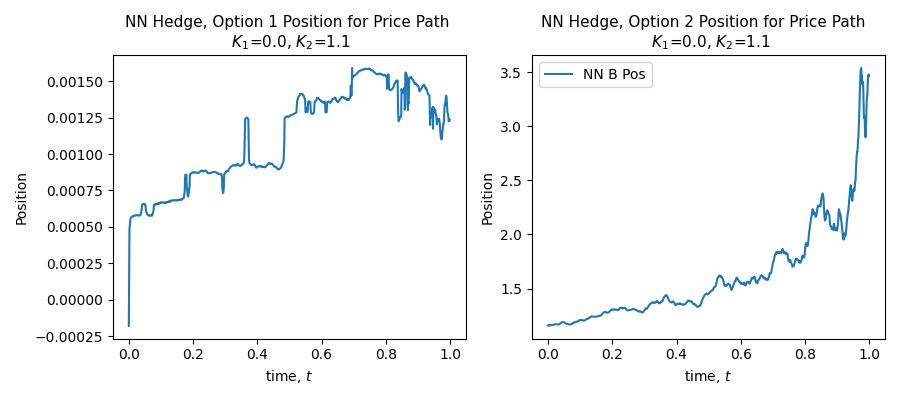}
    \caption{\small{Comparing the gamma hedging method with the strategy chosen by our deep hedging model, applying the $\mathcal{L}^\text{mean}$ loss and high transaction costs.}}
    \label{fig:Abs, Trans - High}
\end{figure}

\begin{figure}[htp!]
\centering
    \includegraphics[width=12cm]{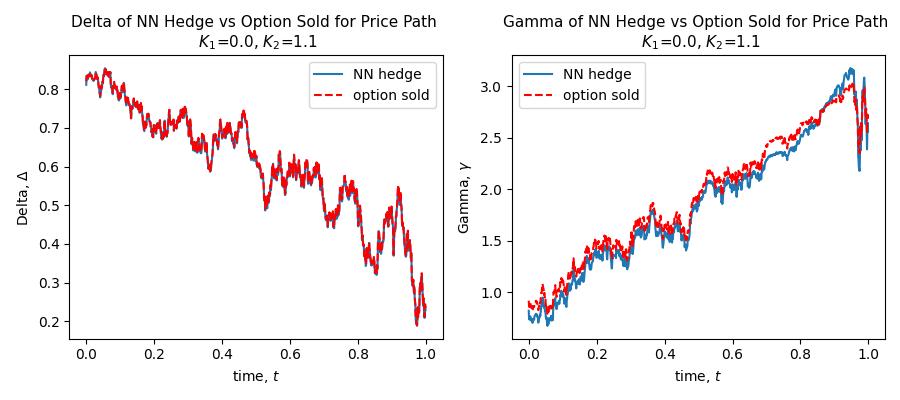}
    \includegraphics[width=12cm]{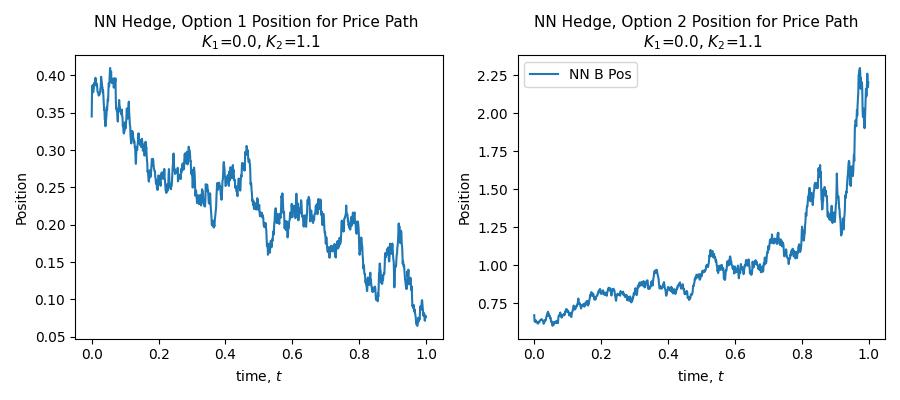}
    \caption{\small{Comparing the gamma hedging method with the strategy chosen by our deep hedging model, applying the $\mathcal{L}^\text{max}$ loss and no transaction costs.}}
    \label{fig:Rob, Trans - No}
\end{figure}

\begin{figure}[htp!]
\centering
    \includegraphics[width=12cm]{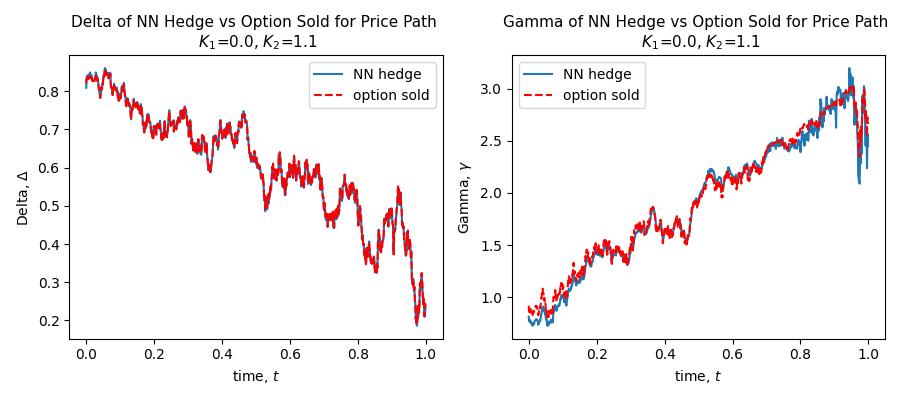}
    \includegraphics[width=12cm]{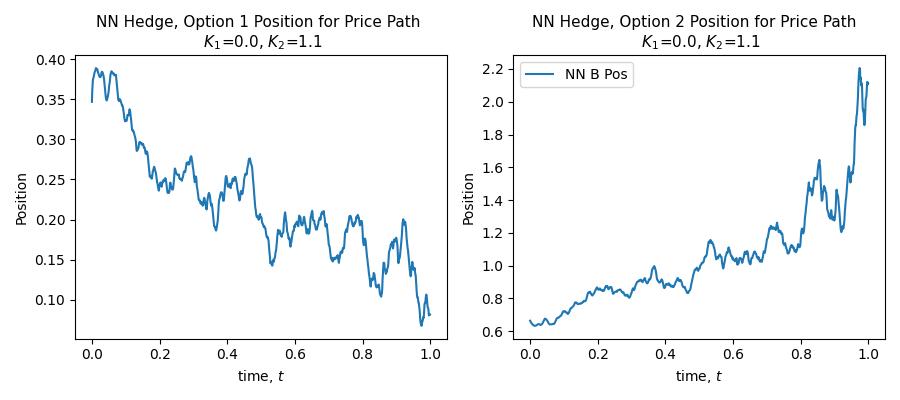}
    \caption{\small{Comparing the gamma hedging method with the strategy chosen by our deep hedging model, applying the $\mathcal{L}^\text{max}$ loss and normal transaction costs.}}
    \label{fig:Rob, Trans - Low}
\end{figure}

\begin{figure}[htp!]
\centering
    \includegraphics[width=12cm]{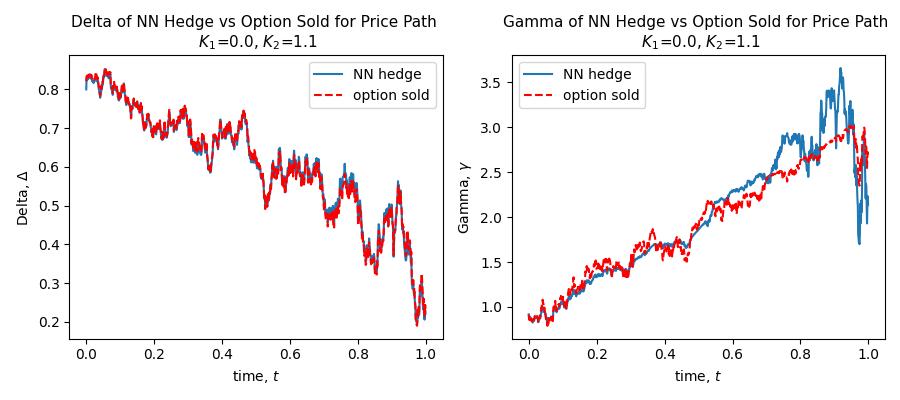}
    \includegraphics[width=12cm]{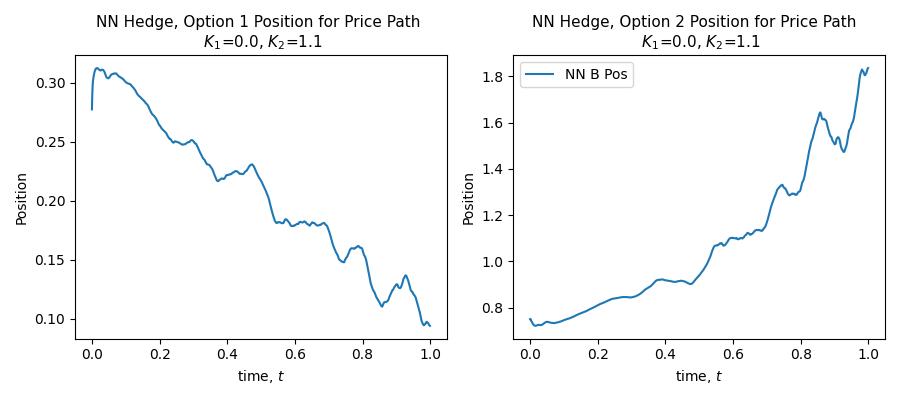}
    \caption{\small{Comparing the gamma hedging method with the strategy chosen by our deep hedging model, applying the $\mathcal{L}^\text{max}$ loss and high transaction costs.}}
    \label{fig:Rob, Trans - High}
\end{figure}

\subsection{Method comparison}\label{Result Sec: Method Comparison}

By simulating $1\times10^5$ underlying price paths under the $\mathcal{L}^\text{mean}$ and $\mathcal{L}^\text{max}$ loss function settings, we can apply the delta and gamma hedging strategies as described in Section \ref{Section: Delta and Gamma Hedging} and derive their performance under the different transaction cost sizes. 

We then add to the models seen in section \ref{Result Sec: Deep hedging with trans} by training models with $M=1$, under each of the given loss function and transaction cost scenarios, and apply these to the same sample as the delta and gamma hedging methods. Table \ref{summary of results} gives the results and allows us to compare the performance of the different strategies. All hedging strategies given use $n=1000$ as in section \ref{Result Sec: Deep hedging with trans}

\newpage

\begin{table}[htp!]
\centering
\begin{tabular}{cp{3.5cm}p{1.8cm}p{1.9cm}p{1.85cm}}
\toprule
\multirow{2}{*}{\textbf{Loss}} &
 \multirow{2}{*}{\textbf{Hedging Method}}  & \multicolumn{3}{c}{\textbf{Transaction Costs}} \\ \cline{3-5} 
         &  & \makecell{$p_1=0\%$\\$p_2=0\%$}
         & \makecell{$p_1=0.005\%$\\$p_2=0.25\%$}
         & \makecell{$p_1=0.5\%$\\$p_2=0.5\%$} \\ \midrule 
 \multirow{4}{*}{$\mathcal{L}^\text{mean}$} & Delta Hedging & $0.41\times 10^{-3}$ & $0.46\times 10^{-3}$ & $20.62\times 10^{-3}$  \\
 & Gamma Hedging & $0.01\times 10^{-3}$ & $2.92\times 10^{-3}$ &  $30.07\times 10^{-3}$ \\ 
 & Deep: 1 Instrument &  $0.63\times 10^{-3}$  &  $0.90\times 10^{-3}$ & $9.12\times 10^{-3}$  \\ 
 & Deep: 2 Instruments & $0.57\times 10^{-3}$  & $0.58\times 10^{-3}$ & $2.83\times 10^{-3}$  \\ \midrule
 \multirow{4}{*}{$\mathcal{L}^\text{max}$} & Delta Hedging &   $6.17\times 10^{-2}$ & $6.27\times 10^{-2}$ & $15.34\times 10^{-2}$ \\
 & Gamma Hedging & $0.03\times 10^{-2}$ & $0.97\times 10^{-2}$ & $8.62\times 10^{-2}$ \\ 
 & Deep: 1 Instrument &  $5.38\times 10^{-2}$   & $5.37\times 10^{-2}$ & $8.38\times 10^{-2}$  \\
 & Deep: 2 Instruments &  $0.49\times 10^{-2}$  & $0.73\times 10^{-2}$  &  $1.60\times 10^{-2}$    \\ \bottomrule
\end{tabular}
\caption{\small{Loss function value for the different hedging models discussed, on a sample of $1.0\times 10^5$ underlying price paths}}
\label{summary of results}
\end{table}

\section{Conclusion}

We have seen that deep hedging has the capacity to learn the optimal hedging approach in the absence of transaction costs. Deep-hedging is effective both for the classical loss function and the loss function incorporating model uncertainty and, as the hedging interval tends to 0, deep learning successfully identifies the theoretical optimum of gamma-hedging.

When transaction costs were introduced, we see that that, as we increase transaction costs, the optimal robust hedging strategy continues to be well-approximated by gamma hedging.

On the other hand, the optimal strategy for minimising mean absolute PnL differs from gamma hedging. Although we cannot clearly describe the strategy chosen, in the case where the same high transaction costs were applied to both Option 1 and 2, it appears the strategy may have been to use Option 2 to maintain the delta hedge rather than the gamma hedge as it would offer cheaper delta when the underlying price is near $K^2$.

Together, these results suggest that in practice gamma hedging is more valuable for its ability to cope with model uncertainty than for reducing transaction costs.

Whilst gamma hedging can offer an improved performance to delta hedging when focusing on robust hedging, once transaction costs are introduce, blindly gamma hedging is suboptimal. Deep-learning can identify a more nuanced adaptation of gamma-hedging which substantially outperforms the other strategies.

Our results demonstrate that when building a hedging portfolio, it is important to consider the many instruments available in the market if one wishes to develop the optimum trading strategy.

\bibliographystyle{alpha}
\bibliography{citations}

\newcommand{\etalchar}[1]{$^{#1}$}
\begin{thebibliography}{WBWB19}

\bibitem[AI23]{armstrong2023gamma}
John Armstrong and Andrei Ionescu.
\newblock Gamma hedging and rough paths, 2023.

\bibitem[BGTW18]{bühler2018deep}
Hans Bühler, Lukas Gonon, Josef Teichmann, and Ben Wood.
\newblock Deep hedging, 2018.

\bibitem[BW09]{broden2009convergence}
Mats Brod{\'e}n and Magnus Wiktorsson.
\newblock {\em On the convergence of higher order hedging schemes}.
\newblock Mathematical Statistics, Centre for Mathematical Sciences, Faculty of~…, 2009.

\bibitem[Car20]{carbonneau2020deephedginglongtermfinancial}
Alexandre Carbonneau.
\newblock Deep hedging of long-term financial derivatives, 2020.

\bibitem[CCHP20]{Cao_2020}
Jay Cao, Jacky Chen, John Hull, and Zissis Poulos.
\newblock Deep hedging of derivatives using reinforcement learning.
\newblock {\em The Journal of Financial Data Science}, 3(1):10–27, December 2020.

\bibitem[CRK{\etalchar{+}}23]{Cherrat2023quantumdeephedging}
El~Amine Cherrat, Snehal Raj, Iordanis Kerenidis, Abhishek Shekhar, Ben Wood, Jon Dee, Shouvanik Chakrabarti, Richard Chen, Dylan Herman, Shaohan Hu, Pierre Minssen, Ruslan Shaydulin, Yue Sun, Romina Yalovetzky, and Marco Pistoia.
\newblock Quantum {D}eep {H}edging.
\newblock {\em {Quantum}}, 7:1191, November 2023.

\bibitem[GM12]{gobet2012tracking}
Emmanuel Gobet and Azmi Makhlouf.
\newblock The tracking error rate of the delta-gamma hedging strategy.
\newblock {\em Mathematical Finance: An International Journal of Mathematics, Statistics and Financial Economics}, 22(2):277--309, 2012.

\bibitem[GWV{\etalchar{+}}23]{Gao_2023}
Kang Gao, Stephen Weston, Perukrishnen Vytelingum, Namid Stillman, Wayne Luk, and Ce~Guo.
\newblock Deeper hedging: A new agent-based model for effective deep hedging.
\newblock In {\em 4th ACM International Conference on AI in Finance}, ICAIF ’23. ACM, November 2023.

\bibitem[HLP94]{RePEc:nbr:nberwo:4718}
James~M. Hutchinson, Andrew Lo, and Tomaso Poggio.
\newblock A nonparametric approach to pricing and hedging derivative securities via learning networks.
\newblock NBER Working Papers 4718, National Bureau of Economic Research, Inc, 1994.

\bibitem[HTZ21]{horvath2021deephedgingroughvolatility}
Blanka Horvath, Josef Teichmann, and Zan Zuric.
\newblock Deep hedging under rough volatility, 2021.

\bibitem[KB17]{kingma2017adam}
Diederik~P. Kingma and Jimmy Ba.
\newblock Adam: A method for stochastic optimization, 2017.

\bibitem[KPKP92]{kloedenPlaten}
Peter~E Kloeden, Eckhard Platen, Peter~E Kloeden, and Eckhard Platen.
\newblock {\em Stochastic differential equations}.
\newblock Springer, 1992.

\bibitem[MWB{\etalchar{+}}22]{murray2022deephedgingcontinuousreinforcement}
Phillip Murray, Ben Wood, Hans Buehler, Magnus Wiese, and Mikko~S. Pakkanen.
\newblock Deep hedging: Continuous reinforcement learning for hedging of general portfolios across multiple risk aversions, 2022.

\bibitem[WBWB19]{wiese2019deephedginglearningsimulate}
Magnus Wiese, Lianjun Bai, Ben Wood, and Hans Buehler.
\newblock Deep hedging: Learning to simulate equity option markets, 2019.

\end{thebibliography}

\newpage

\end{document}